\documentclass{rspublic}
\usepackage{amsmath}
\usepackage{amsfonts}
\usepackage{amssymb}
\usepackage[Symbol]{upgreek}
\usepackage[nointegrals]{wasysym}
\usepackage{mathrsfs}
\usepackage{graphicx}
\newcommand{\real}{{\mathbb R}}
\newcommand{\bfu}{{\bf u}}
\newcommand{\bfv}{{\bf v}}
\newcommand{\bfw}{{\bf w}}
\newcommand{\bfx}{{\bf x}}
\newcommand{\bfy}{{\bf y}}
\newcommand{\bfa}{{\bf a}}
\newcommand{\bfb}{{\bf b}}
\newcommand{\bfr}{{\bf r}}
\newcommand{\bfe}{\hat{\bf e}}

\newcommand\pbrk[1]{\left(#1\right)}
\begin{document}

\title[Navier-Stokes and Euler Equations]{\large Leray self-similarity equations in fluid dynamics}

\author[F. Lam]{F. Lam}

\affiliation{ }

\label{firstpage}

\maketitle

\begin{abstract}{Euler Equations; Navier-Stokes Equations; Vorticity Equation; Finite Energy; Leray Self-similar Equation; Singularity; Viscosity; \\ Lebesgue Norms; Sobolev Inequality}

In the present note, we show that, as {\it a priori} bounds, the vorticity dynamics derived from Leray's backward self-similarity hypothesis admits only trivial solution in viscous as well as inviscid flows. By analogy, there is no non-zero solution in the forward self-similar equation. Since the Navier-Stokes or Euler equations are invariant under space translation in the whole space, our analysis establishes that technically flawed arguments have been exploited in a number of past papers, notably in Ne\v{c}as, R\r{u}\u{z}i\u{c}ka \& \u{S}ver\'{a}k (1996); Tsai (1998); and Pomeau (2016), where the presumed decays or bounds at infinity are ill-defined and non-existent. Furthermore, an effort has been made to exemplify an inappropriate application of the familiar extremum principles in the theory of linear elliptic equation.

In the appendix, we give a counterexample to the Sobolev inequality and, hence illustrate the nature of self contradiction. In the totality comparison of $L_p$ norms, its scope of application is not significant.
\end{abstract}
\section{Introduction}
In real fluid of uniform density $\rho$, the Navier-Stokes equations of motion are derived from Newton's second law of motion, and mass conservation
\begin{equation} \label{ns}
	\partial_t \bfu + (\bfu . \nabla) \bfu  = - {\rho}^{-1} \nabla p + \nu \Delta \bfu,\;\;\; \nabla.\bfu=0, 
\end{equation}
where the velocity vector $\bfu(\bfx,t)=(u,v,w)(\bfx,t)$, the coordinates $\bfx=(x_1,x_2,x_3)$, and the scalar $p(\bfx,t)$ denotes the pressure, $\nu$ the kinematic viscosity. The scalar pressure can be eliminated from the system of equations (\ref{ns}) 
\begin{equation} \label{vort}
	\partial_t \upomega - \nu \Delta \upomega = (\upomega . \nabla) \bfu  - (\bfu . \nabla )\upomega,
\end{equation}
where the vorticity field, $\upomega{=}\nabla {\times} \bfu$, inherits the velocity solenoidal property $\nabla.\upomega{=}0$. 
As a consistency check in practice, the continuity enforces the compatibility
\begin{equation} \label{cmp}
\Delta \bfu = - \nabla{\times}\upomega.
\end{equation} 
This is a direct consequence of vector identity $\nabla{\times}(\nabla{\times}A)=\nabla(\nabla.A)-\nabla.(\nabla A).$ Alternatively, the momentum equation of (\ref{ns}) can be re-written as
\begin{equation} \label{ns1}
	\partial_t \bfu + \upomega{\times} \bfu  = - \nabla (p/\rho+|\bfu|^2/2) + \nu \Delta \bfu.
\end{equation}

We are interested in finite-energy initial value problems subject to the initial data
\begin{equation} \label{ic}
 \bfu(\bfx,0)=\bfu_0(\bfx)\; \in \; C_c^{\infty}(\real^3), 
\end{equation} 
where the initial solenoidal velocity is assumed to be smooth with compact support. 

Suppose that there be a fluid with zero viscosity, the momentum equation reduces to the Euler equation by formally setting $\nu=0$
\begin{equation} \label{euler}
	 \partial_t \bfu  + (\bfu . \nabla) \bfu = - {\rho}^{-1} \nabla p.
\end{equation}

The adjoints of the governing equations are given by
\begin{equation} \label{nsadj}
	\partial_t \bfu^{\dagger} + \nu \Delta \bfu^{\dagger} = - (\bfu^{\dagger} . \nabla) \bfu^{\dagger}  - {\rho}^{-1} \nabla p^{\dagger},\;\;\; \nabla.\bfu^{\dagger}=0. 
\end{equation}
In particular, the fundamental solution of the backward diffusion operator,
\begin{equation*}
H(\bfx,\bfy,t,s)={\big(\:4 \pi \nu (s{-}t)\:\big)^{-3/2}} \exp \Big(- \frac{|\bfx-\bfy|^2}{\;4 \nu (s-t)\;}\;\Big),\;\;\; t<s,\;\;(\bfx,\bfy) \in \real^3,
\end{equation*}
implies the irreversibility of viscous flows, in contrast with the non-viscous flows governed by the Euler equations. The equations in (\ref{ns}) constitute a parabolic system of equations, and (\ref{euler}) a hyperbolic one. We observe that the non-linearity keeps its identity under time reversal.

Taking divergence of (\ref{ns1}) or (\ref{euler}) and using the continuity, we obtain the Poisson equation for the pressure 
\begin{equation} \label{ppoi}
	\Delta p(\bfx)= - \rho \: \big(\: |\nabla \bfu|^2 - |\nabla{\times}\bfu|^2  \:\big)(\bfx),
\end{equation}
which holds at every instant of time $t$.

One of the important properties of the Navier-Stokes or Euler equations in $\real^3$ is that they are invariant to space translation
\begin{equation} \label{inv}
\bfu(\bfx,t)=\bfu(\bfx-\bfa,t),\;\;\;p(\bfx,t)=p(\bfx-\bfa,t),\;\;\;\upomega(\bfx,t)=\upomega(\bfx-\bfa,t),
\end{equation}
where bold-face symbol $\bfa=(a_1,a_2,a_3)$ denotes an arbitrary finite vector. The same principle applies to the adjoints. It is easy to understand the invariance, because the equations of motion do not contain explicit terms of the co-ordinates.
\section{Leray's similarity solutions}
\subsection*{Backward self-similarity}
In view of the translation invariant principle, Leray's proposal (1934) of transforming the equations of motion takes the form 
\begin{equation} \label{leray}
\begin{split}
 \bfu(\bfx-\bfa,t) &= {t^*}\; \bfv ( (\bfx-\bfa) \; t^*)={t^*}\; \bfv(\bfy-\bfb),\\
 p(\bfx-\bfa,t) &= {{t^*}^2}\; q( (\bfx-\bfa) \; t^*)={{t^*}^2}\; q(\bfy-\bfb),
\end{split}
\end{equation}
where 
\begin{equation*}
t^*=\frac{1}{\sqrt{\:2\sigma(T_s-t) \:}}
\end{equation*}
for constant $\sigma > 0$, the symbol, $T_s(\: \geq t > 0)$, stands for a potential singularity time. Effectively, the stretched co-ordinates, $(\bfx-\bfa) t^*=\bfy-\bfb$, become time-independent for arbitrary vector $\bfb=(b_1,b_2,b_3)$. Because $t^*$ is assumed to be as large as we may please near the singular time, the Leray reduction is meaningful only in the whole space $\real^3$. We find (for continuous function $f$)
\begin{equation} \label{tm}
\frac{\partial t^*}{\partial t}=\sigma {t^*}^3,\;\;\;t^*\frac{\partial f}{\partial t^*}=t^*\frac{\partial y_i}{\partial t^*} \frac{\partial f}{\partial y_i}=(x_i-a_i)t^* \frac{\partial f}{\partial y_i}=(y_i-b_i) \frac{\partial f}{\partial y_i},
\end{equation}
where the last identity underscores the essence of Leray's time deletion, namely the co-ordinates space is {\it uniform} in time-contraction or dilation, $(y_i - b_i)\leftrightarrows (x_i-a_i) \: t^*$. 
Making use of identities (\ref{tm}), equations (\ref{ns}) are transformed into an elliptic system (cf. Leray's (3.11))
\begin{equation} \label{le}
	\sigma \big( \bfv + ((\bfy-\bfb). \nabla) \bfv \big) + (\bfv.\nabla)\bfv=-\nabla q/\rho + \nu \Delta \bfv,\;\;\; \nabla. \bfv=0,
\end{equation}
where differentiations are with respect to $\bfy$. For initial data of finite energy, $\bfu_0$, the transformed `initial' velocity is given by 
\begin{equation*}
\bfv_0(\bfy-\bfb)=\sqrt{2\sigma T_s}\: \bfu_0(\bfx-\bfa),
\end{equation*}
which is localised and translation-invariant because $\bfu_0$ is. At least, the size of the transformed velocity $\bfv$ defined in the elliptic equation is comparable to that of $\bfu$. However, the introduction of the parameter $\bfb$ does not affect the driving force in the pressure $q$-Poisson equation which has the identical form of (\ref{ppoi}), as $\nabla.(((\bfy-\bfb).\nabla)\bfv)=0$.

Denote the vorticity in the dilated space by 
\begin{equation*}
\bfw(\bfy)=\nabla{\times}\bfv(\bfy)={t^*}^{-2}\:\upomega(\bfx),
\end{equation*}
which is also solenoidal, having components $\bfw=(w_1,w_2,w_3)$. The vorticity description of the transformed momentum (\ref{le}) is written as 
\begin{equation} \label{evt}
\sigma\big( 2\bfw + ( \:(\bfy-\bfb). \nabla ) \bfw \big) -\nu \Delta \bfw= (\bfw.\nabla)\bfv-(\bfv.\nabla)\bfw.
\end{equation}
In the present note, we intend to solve the system (\ref{le}) or the vorticity equation (\ref{evt}) by first establishing the {\it a priori} bounds.
\subsection*{Implications of Leray reduction}  
The Leray transform (\ref{leray}) implies that the functions, $\bfu$ and $\bfv$, must be regular functions of space and time ($t < T_s$). Certainly, they do not belong to the equivalent Lebesgue class $L^s(\real^3)$, for $1 \leq s < 3$ (say), as it makes no sense to assign a numerical value to the velocity at a given $t^* \bfx$, if $\bfu, \bfv \in L^2(\real^3)$, because the sets of measure zero in $\bfu$ and $\bfv$ do not overlap in general. 

For the singularity at $t=T_s$, velocity $\bfv$ has to be twice differentiable in space up to $T_s$, at least, in order for the dynamic equations to be meaningful. Leray's conjecture is that $\bfu$ may become arbitrarily large at the point $t=T_s$ if (\ref{le}) admits a non-zero regular solution $\bfv$. Thus, the issue of the self-similar blow-up is directly related to the regularity of the Navier-Stokes equations.

Bafflingly, we notice that the non-linearity has been left unscathed in (\ref{le}) and, inevitably, in (\ref{evt}). In other words, the conversion of the dynamics to a quasi-steady problem by the space-inflation has not really addressed the key issues. The second point to notice is that the mass conservation in (\ref{le}) is valid only for $0\leq  t < T_s$ as we obtain it from 
\begin{equation*}
{\partial u_i}/{\partial x_i}= {t^*}^2 \: {\partial v_i}/{\partial y_i}=0,
\end{equation*}
with the assumption of $t^*$ being finite. At the singular time, $t^* \rightarrow \infty$, the principle of mass conservation breaks down entirely. What happens to the material of fluid? This scenario of absurdity is absolutely irrational. Lastly, the momentum equation has a definite meaningful only if ${t^*}^3$ is non-zero but finite. The implication is, at the instant of singularity, the laws of classical mechanics underlying incompressible fluid dynamics must fail. 

On the other hand, the energy and enstrophy are found to be
\begin{equation*}
\int_{\real^3} \bfu^2(\bfx,t)\; \rd \bfx = \; \frac{1}{t^*}\int_{\real^3} \bfv^2(\bfy,t^*) \;\rd \bfy,
\end{equation*}
and
\begin{equation*}
\int_{\real^3} \upomega^2(\bfx,t)\; \rd \bfx = \; t^*\int_{\real^3} \bfw^2(\bfy,t^*) \; \rd \bfy ,
\end{equation*}
respectively. At the assumed $t \rightarrow T_s$, the energy disappears instantly or the complete flow halts suddenly, while the enstrophy surges without bounds. This is a peculiar `physical' phenomenon. Nevertheless, the accumulative dissipation is given by
\begin{equation*}
\nu \int_0^{T_s-\varepsilon} \!\!\! \int_{\real^3} \upomega^2(\bfx,t)\; \rd \bfx \rd t \leq  \; \nu A \sqrt{2\:(T_s-\varepsilon)/\sigma} < \infty
\end{equation*}
where $A=\sup_{0\leq t < T_s-\varepsilon} \|\bfw\|_{L^2(\real^3)}$, because the singularity is integrable. As far as the law of energy conservation is concerned, the role of Leray's conjecture on singularity is rather obscured. At least, its presence does not conform to the observation of numerous turbulent flows. A practical example is the turbulent flow at the edge of a boundary layer, where vorticity eddies, large or small, are engaged in highly unsteady entrainment processes which occur over periods of finite time. In many daily applications, when $t^* \sim O(100)$ or at moments when $t$ is sufficiently near $T_s$, it may not be appropriate to treat velocity $\bfu$ in (\ref{leray}) as incompressible, in comparison to the order of $\bfv_0$.
\subsection*{Solution of Leray equation}
Let the total quantities, $W=w_1+w_2+w_3$, and $V=v_1+v_2+v_3$. By summing the vectors (\ref{evt}), we obtain a scalar equation,
\begin{equation} \label{etvt}
\sigma \big( 2W +(\:(\bfy-\bfb). \nabla ) W \big) - \nu \Delta W = (\bfw.\nabla)V - (\bfv.\nabla)W,
\end{equation}
which is much simpler to analyse. Integrating over the whole space and making use the solenoidal property in $\bfv$ and $\bfw$, we derive the following integral discriminant:
\begin{equation} \label{qvt}
\begin{split}
\int_{\real^3} \big(\: W {+} {\nu}\Delta W/\sigma\:\big)\: \rd \bfy = &\int_{\real^2} (y_1{-}b_1) ( w_1+w_2+w_3 ) \: \Big|_{|y_1|{\rightarrow} \infty} \rd y_2 \rd y_3 \\
\quad & \hspace{3mm} + \int_{\real^2} (y_2{-}b_2) ( w_1+w_2+w_3 ) \: \Big|_{|y_2|{\rightarrow} \infty} \rd y_3 \rd y_1 \\
\quad & \hspace{9mm} + \int_{\real^2} (y_3{-}b_3) ( w_1+w_2+w_3 ) \: \Big|_{|y_3|{\rightarrow} \infty} \rd y_1 \rd y_2,
\end{split}
\end{equation}
because a cancellation due to symmetry renders the non-linearity on the right of (\ref{etvt}) to zero. 
Since each of the nine boundary terms involving $w_i$ must be independent of the co-ordinates dilation or contraction or the choice of the arbitrary vector $\bfb$, the logical deduction is that 
\begin{equation} \label{iqvt}
w_1 = w_2 = w_3 = 0.
\end{equation}
As a consistent outcome, the left-hand integral vanishes identically for any $\nu$ and $\sigma$. We would like to emphasise the fact that we may still deduce the null vorticity condition if we replace $\bfy-\bfb$ by $(\bfx-\bfa) t^*$ on the right-hand side of (\ref{qvt}) and let $|\bfx|$ become large. Thus the trivial vorticity state also applies to the original space-time co-ordinates $(\bfx,t)$, regardless of the time dilation. 

To establish the differentiable properties, we take differentiations $\partial^{\alpha}_{y_j}$ (for multi-index $\alpha \geq 1$, $\alpha$ being a natural number) on equation (\ref{etvt}). We then integrate the resulting equations. The important step is on the non-linear terms 
\begin{equation*}
\int_{\real^3} \partial^{\alpha}_{y_j} \Big(\: (\bfw.\nabla)V - (\bfv.\nabla)W \: \Big) \rd \bfy =  \partial^{\alpha}_{y_j} \int_{\real^3}\Big( \: (\bfw.\nabla)V - (\bfv.\nabla)W \: \Big)\rd \bfy. 
\end{equation*}
In other words, the differentiations and integration on the non-linearity commute. This remarkable property enables us to derive
\begin{equation*} 
\int_{\real^3} \big(\partial^{\alpha}_{y_j} W\big) \: \rd \bfy,
\end{equation*}
in terms of boundary data similar to those in (\ref{qvt}). We do not wish to repeat every detail of the mechanical operations. For instance, the revised first term on the right of (\ref{qvt}) now reads, for $\alpha>1$,
\begin{equation*}
\int_{\real^2} (y_1-b_1) \partial^{\alpha}_{y_j} (w_1+w_2+w_3) \: \Big|_{|y_1|{\rightarrow}\infty} \rd y_2 \rd y_3.
\end{equation*}
For $\alpha=1$, consider $\partial_{y_1}W$, the analogous first term is given by
\begin{equation*}
\int_{\real^2} \Big( (y_1-b_1) \frac{\partial W}{\partial y_1} + W \Big) \: \Big|_{|y_1|{\rightarrow}\infty} \rd y_2 \rd y_3.
\end{equation*}
Consequently, the following derivative nullification must hold
\begin{equation} \label{iqvtd}
\partial^{\alpha}_{y_j} w_i = 0.
\end{equation}

By the incompressibility, $\Delta \bfv = - \nabla{\times} \bfw =0$, we derive the smoothness properties of the velocity field from the identities, $\Delta \partial^{\alpha}_{y_j} v_i = - (\nabla{\times} \partial^{\alpha}_{y_j} \bfw )_i=0$ for every $i,j=1,2,3$. In view of Liouville's theorem for harmonic functions, we know that each component of $\bfv$ is a constant. However, we must carry out additional analysis to verify if it is zero. Under the translation by $\bfb$, `large distances' are not properly defined; any decay rate at infinity for the velocities or their derivatives do not make sense. One way forward is to restore to dynamics, assuming no knowledge of $\bfv$. We first write the momentum in the form
\begin{equation*}
\frac{1}{\rho}\frac{\partial q}{\partial y_i} = - \sigma \: \Big(\: (y_i-b_i)\frac{\partial v_i}{\partial y_i} + v_i \: \Big) + F_i(\bfv,\sigma),
\end{equation*}
where, evidently, every $F_i$ vanishes if each component of $\bfv$ does.
Integrating with respect to $y_i$, the three expressions for the pressure are given by
\begin{equation*}
q/\rho= q_{\bfr}-\sigma (y_i-b_i)v_i \Big|_{r_i}+ \int_{r_i}F_i(\bfv,\sigma) \: \rd y_i,
\end{equation*}
where $q_{\bfr}$ stands for the gauge pressure at the reference location $\bfr=(r_1,r_2,r_3)$. To avoid a contradiction on $q$ for arbitrary $\bfb$, we assert that $v_1=v_2=v_3=0$. To be definite, we choose the zero gradient, $\nabla q=0$.  

By the same token, our arguments leading to non-existence apply equally well to ideal fluids because setting $\nu \equiv 0$ formerly in the integration condition (\ref{qvt}) does not change the right-hand discriminant. In summary, the Leary blow-up scenario never occurs in inviscid incompressible flows. 

Strictly speaking, the pressure in the Leray transform always entails blows up at $t=T_s$ as the pressure $q$ is only determined up to an arbitrary constant. Dynamically, the gradients of a singular pressure cause imbalance of the momenta so that the flow field evolves irregularly, from the very beginning. To avoid this embarrassment, the pressure in (\ref{leray}) must be replaced by
\begin{equation} \label{leray2}
\nabla p(\bfx-\bfa,t)={t^*}^3 \:\nabla q(\bfy-\bfb),
\end{equation}
where the gradient operation on each side is clear.

To understand why the Leray equations do not admit non-trivial solutions, we first determine the adjoints of $\bfv$,
\begin{equation} \label{adj}
	\sigma \bfv^{\dagger} - \sigma \big((\bfy-\bar{\bfb}). \nabla \big) \bfv^{\dagger} - \nabla q^{\dagger}/\rho - (\bfv^{\dagger}.\nabla)\bfv^{\dagger}- \nu \Delta \bfv^{\dagger}=0,\;\;\; \nabla. \bfv^{\dagger}=0,
\end{equation}
where $\bar{\bfb}=\bfb-\bfe$, and $\bfe$ is unit vector. These equations are a consequence of the Lagrange-Green identity for differential operators. Compared to Leray's original equations (\ref{le}), we notice that the criticality of the non-linear term has hardly been stressed, apart from some changes in the sign of the coefficients. However, the second term on the left shows the original dynamics $\partial_t \bfu$ is further stretched. Equations (\ref{adj}) can be analysed and solved by the same method of the total vorticity. In brief, similar to the discriminant condition (\ref{qvt}), the modified left-hand integrand is $5W^{\dagger}{-}\nu \Delta W^{\dagger}/\sigma$. The conclusion of trivial adjoint solutions, $\bfv^{\dagger}=0, \nabla q^{\dagger}=0$, follows as {\it a priori} bounds. 

It is instructive to elucidate the triviality result of the Leray velocity field in terms of physics. The equations of motion have a mathematical solution, 
\begin{equation*}
\bfu(\bfx,t)={\mbox{const}},\;\;\; p(\bfx,t) = {\mbox{const}}.
\end{equation*}
In the case of the whole space, we do not know how to initiate this constant motion as it contains infinite energy. Physically, a Galilean transform renders the constant flow to the stationary state. In our finite-energy flow (\ref{ic}), developed velocity $\bfu$ cannot be everywhere non-zero at all times. Thus, it is plausible and necessary that there are multiple stagnation points, or possibly stagnant flow regions, where $\bfu=0$, throughout the flow development. Let us denote any one of these instantaneous stagnations by $\bfu_{sp}(\bfx_{sp},t)$. Consider any instant, $t=t_1$, prior to a potential singularity at a location $\bfx_s$ when $T_s-t_1 \neq 0$. We then have $\bfv(t^*\bfx_s)=\bfu(\bfx_s,t_1)\:{\sqrt{2 \sigma (T_s-t_1)}}<\infty$. The transform, $t^* \bfx$, in fact stretches the co-ordinates, or simply, it sends $\bfx_s$ to the location far away to $\bfx^*$ (say). As the $\real^3$-space is homogeneous, we can equate $\bfu(\bfx_s,t_1)$ to $\bfu(\bfx_{sp},t_1)$ by translating $\bfx_s$ to $\bfx_{sp}$ which, effectively, offsets the Leray reduction. It follows that the value $\bfv(\bfx^*)$ must coincide with $\bfu_{sp}(\bfx_{sp},t_1)$ as $\bfx_s \rightarrow \bfx_{sp}$. Fundamentally, Leray's dilation (\ref{leray}) does not constitute a dynamic transformation on fluid motions.
\subsection*{Forward self-similar decaying solution}
If the time dilation $t^*$ is replaced by temporal monotonic decay
\begin{equation*}
\tilde{t^*}=\frac{1}{\sqrt{\:2\sigma(t-T) \:}},\;\;\;\;\;\;t>T\geq0,
\end{equation*}
in the reduction (\ref{leray}), the resulting equations describe forward self-similar flows $(\tilde{\bfv},\tilde{q})$. Then system (\ref{le}) now reads
\begin{equation} \label{fle}
	-\sigma \big( \tilde{\bfv} + ((\bfy-\bfb). \nabla) \tilde{\bfv} \big) + (\tilde{\bfv}.\nabla)\tilde{\bfv}=-\nabla \tilde{q}/\rho + \nu \Delta \tilde{\bfv},\;\;\; \nabla. \tilde{\bfv}=0.
\end{equation}
Essentially, the analytical structure remains unchanged compared to (\ref{evt}). The only difference is that there is a sign change in the parameter $\sigma$, i.e., $t^*(-\sigma)=\tilde{t^*}(\sigma)$. We only need to apply a negative sign to $\sigma$ in the forward vorticity (the counterpart of (\ref{etvt})), and in the left-side integral (\ref{qvt}). Without further ado, we conclude that $(\tilde{\bfv}=0, \:\nabla\tilde{q}=0)$ for $\nu \geq 0$.
\section{Recap of related past work}
In the introduction of the self-similar solutions (\ref{le}) or (\ref{fle}), Leray (see \S 20 of his 1934 paper) mentioned that he was unsuccessful in studying these equations. We do not know for sure whether he was unable to establish non-zero solutions for $\bfv$. Perhaps, he was aware of the fact that his equations had only zero solution. Nevertheless, the introduction of transform (\ref{leray}) implies the existence of regular solutions for both $\bfu$ and $\bfv$, at least over $t<T_s$. The switch to smooth functions by Leray for possible point-wise singularities deviated considerably from his central idea of generalised solutions to the Navier-Stokes equations, where `turbulent solutions' were described by weakly differentiable functions.

The paper by Rosen (1970) is difficult to follow, as it contains obvious analytic errors, as far as the similarity solutions are concerned. It appears that he was the first to state that the Leray equations admit trivial solution. Foias \& Temam (1983) made an attempt to investigate homogeneous turbulent flows described the Leray equation (\ref{le}) in a cube where periodic boundary conditions or symmetries must have been assumed. They concluded that, in the remark on p.200, Leray's reduction does not admit non-zero steady homogeneous statistical solutions, satisfying a regularity condition, and a growth constraint. We shall not discuss their work further in the present note because the complete unsteady Navier-Stokes equations are ill-formulated in the periodic setting. In addition, there is no justifiable rationale to treat turbulent flows as stationary by ignoring the contribution from $\partial_t \bfu$. 

In the regularity theory of incompressible viscous flows, it is a popular opinion that the mathematical demonstration for the nonexistence of the backward self-similar solutions has been settled in the work of Ne\v{c}as {\it et al.} (1996); Tsai (1998); M\'alek {\it et al.} (1999). However, we believe that this view is barely complete, as it overlooks the fact that these papers contain technical errors, contradicting the properties of the governing equations. We shall be brief and concise so as to highlight the underlying principles. 
\subsection*{Non-existence of large-distance decay}
First, as a principle, the regularity specification in the transform (\ref{leray}) does not imply localisation. For instance, as a standard procedure, we may multiply (\ref{le}) by $\bfv$ and integrate the result to obtain the following energy law (setting $\bfb=0$ temporarily)
\begin{equation*}
\frac{\sigma}{2}\int_{\real^3} |\bfv|^2 \rd \bfy - \nu \int_{\real^3} |\nabla \bfv|^2 \rd \bfy=0,
\end{equation*}
with the assumption,
$(v_1+v_2+v_3) |\bfv|^2 |_{|\bfy|\rightarrow \infty} \rightarrow 0$,
which is entirely unjustified without {\it a priori} bounds, because we are working on a potential singularity for $\bfu$ where the decay implicitly forces certain regularity. In fact, velocity decays of similar nature cannot exist because of the space-invariance (\ref{inv}). Consider, for example, the following decay
\begin{equation*}
|\nabla \bfv | \sim O \big(\: |\bfy|^{-3} \:\big),\;\;\;\;|\bfy| \rightarrow \infty,
\end{equation*}
as stated in Lemma $3.2$ of Ne\v{c}as {\it et al.} (1996). This decay surges into a divergence if the parameter $\bfb$ translates the gradient to any close vicinity of the `large distance'. Similarly, the claim made in Tsai's Lemma $3.1$ (1998), 
\begin{equation*}
\|\nabla \bfv \|_{L^2(B)} \sim o \big(\: \sqrt{\:|\bfy_0|\:} \: \big)\;\;\;\;\;\;\mbox{as}\;\;\;\;\;\; |\bfy_0| \rightarrow \infty,
\end{equation*}
in the unit ball $B$ centred at location $\bfy_0 \in \real^3$, is misleading. By applying a suitable co-ordinates shift on the centre $\bfy_0 \rightarrow \bfy_0{-}\bfb$, the local enstrophy density $\|\nabla \bfv\|_{L^2}$ of the ball is degenerated into a negligible quantity for $|\bfy_0-\bfb| \approx 0^+$. Moreover, the relations of local energy to fluid motions explained by Tsai (1998) make no sense. In essence, the space-invariant law states that decays for $\bfv$ and $q$ at infinity or in confined locality have no precise meanings.  In \S3 of M\'alek {\it et al.} (1999), there were a number of manipulations or repeated assumptions of `local' distances $|\bfy|$, whose sizes are vague according to the invariant principle $(\ref{inv})$. The approach of these papers squarely hinges on their decay rates or large-distance bounds which are ill-founded in $\real^3$ topology. Their proposed proofs of the Leray equations therefore contain illogical inference.

The discussions on the forward self-similar solutions given in Giga \& Miyakawa (1989), and Okamoto (1997), are analytically unsound, as these authors assume decays at infinity, or explicitly,
\begin{equation*}
|\tilde{\bfv}| \sim o(\:|\bfy|^{-3/2}\:),\;\;\;{\mbox{and}},\;\;\; |\tilde{q}| \sim O(\:|\bfy|^{-1/2}\:).
\end{equation*}
Both estimates collapse under the linear shift $\bfv \rightarrow \bfv {-} \bfb \approx 0$ by appropriate choices of $\bfb$. The arguments of Giga \& Miyakawa (1989) were not completely self-consistent. They operated on the Biot-Savart law, or more generally our incompressibility relation (\ref{cmp}) as functions of time, see their equations (1.3), (4.4), and iteration scheme (4.8).\footnote{We encounter this type of `popular' mistakes in incompressible Navier-Stokes literature: integrating or differentiating elliptic equations (of space variables) with respect to time so as to obtain time-dependent bounds, e.g., $L^p(t)$ norms. The pressure Poisson in Ne\v{c}as {\it et al}. (1996) was falsely treated to depend on time; those bounds given in the displayed equation beneath their (3.8) are fictitious. As a further example, Serrin (1962) derived his time-wise derivatives from the velocity-vorticity recovery integral, i.e., the integral form of our (\ref{cmp}), see his formula (20) or (22). Similar miscalculations are found in planar domains, see, for instance, equation (2.29) of Ben-Artzi (1994).}
\subsection*{Misused maximum principle}
Second, the following quantity,
\begin{equation} \label{chi}
\chi=|\bfv|^2/2 + q/\rho + \sigma \bfy. \bfv,
\end{equation}
plays the key role in the Leray solution for the backward similar flows, see Ne\v{c}as {\it et al.} (1996); Tsai (1998). The first two terms are the Bernoulli-Euler pressure, giving an impression that the Bernoulli theorem of potential flow has been generalised for viscous flows. Intuitively, such an extension sounds too good to be true. In fact, under the co-ordinates translation, 
(\ref{chi}) now reads
\begin{equation} \label{chi2}
\chi^*=|\bfv|^2/2 + q/\rho + \sigma (\bfy-\bfb). \bfv,
\end{equation}
which varies over space according to the location $\bfb$. This casts doubt on whether the target function $\chi$ can ever be monotonic. Following Ne\v{c}as {\it et al.} (1996), it is straightforward to shown that $\chi$ satisfies 
\begin{equation} \label{ntm}
 \nu \Delta \chi - \big((\bfv + \sigma \bfy ). \nabla \big)\chi = \nu |\bfw|^2,
\end{equation}
which is an algebraic identity, where the equality sign strictly holds for any flow solution. No explicit functional forms, such as $\bfv=f(\bfy)$, have been given by these authors; the velocity and vorticity fields have remained unknown. A direct observation suggests that the maximum (or minimum) principle for linear equations (see, for instance, Chapter 2 of Protter \& Weinberger 1984; Chapter 3 of Gilbarg \& Trudinger 1998) {\it does not} apply to the function $\chi$, because the left-side functional (\ref{ntm}) is {\it non-linear}, as the coefficient itself must depend on $\chi$. The reason is that the pressure component $q$ is governed by its Poisson equation (cf. an analogous equation to (\ref{ppoi})), and hence it must be a function of $\bfv$, namely $q=q(\bfv(\bfy))$. It follows that the target $\chi=\chi(\bfv(\bfy))$ from (\ref{chi}). By the theory of implicit or inverse functions, we assert that $\bfv=\bfv(\chi(\bfy))$. The exception is, {\it a posteriori}, the trivial state, $\bfv=0$, and $q=\const$ In no circumstances, can the exact equality in (\ref{ntm}) be extended to an inequality, as done by Ne\v{c}as {\it et al.} (1996); Tsai (1998); and M\'alek {\it et al.} (1999). It is not clear if the mistranslation of the non-linearity (\ref{ntm}) inherits a flawed proposition given in Lemma 2.4 of Gilbarg \& Weinberger (1974).  

In general, the size of coefficient function $\bfv + \sigma (\bfy-\bfb)$ in (\ref{chi2}) may substantially vary compared to the principal term $\nu \Delta \chi^*$ due to the presence of the freely-chosen vector $\bfb$. Since we have disproved any sufficiently rapid decays for $\bfv$ or $q$, the coefficient can be made arbitrarily large so as to violate the limiting condition of linear elliptic equations, see, for instance, constraint (3.3) of Gilbarg \& Trudinger (1998). 
\subsection*{Remarks on Serrin's Bernoulli theorem}
There is a parallel between (\ref{ntm}) and Serrin's claim of a maximum principle for steady flows, see \S 76 of Serrin (1959). For the sake of discussion, his equation (76.4) is reproduced
\begin{equation} \label{smx}
\nu \Delta \bar{\chi} - (\bfu.\nabla)\bar{\chi} = \nu |\nabla{\times}\bfu|^2, 
\end{equation}
where $\bar{\chi}=|\bfu|^2/2+p/\rho$ is the Bernoulli-Euler pressure. Specifically, $\bfu=\bfu(\bar{\chi})$ so that the middle term is non-linear in general. The steady momentum equations give the formula
\begin{equation*}
(\bfu.\nabla)\bar{\chi}= \nu (\bfu.\Delta)\bfu.
\end{equation*}
By the identity 
\begin{equation*}
\Delta (|\bfu|^2/2)=(\bfu.\Delta)\bfu + |\nabla \bfu|^2,
\end{equation*}
and the pressure Poisson (\ref{ppoi}), we see that the left-hand side of (\ref{smx}) is reduced to the vorticity norm on the right. Thus the equation is nothing more than another derived expression of the formulas for the energy and pressure, which is exactly satisfied by any steady Navier-Stokes solution.

We voice our objections to the superficial treatment of the non-linear term, because it amounts to mitigating the difficulties of fluid dynamics. It is well-known that there exists a vector potential $\psi(\bfx,t)$ such that $\bfu=\nabla{\times}\psi$. Following the logic to arrive (\ref{ntm}) by simply grouping $(\bfv,q)$ into $\chi$, can we `interpret' the vorticity as
\begin{equation*}
\upomega=\nabla{\times}\bfu=\upomega(\bfu)=\upomega(\psi)=\upomega(\bfx,t),
\end{equation*}
so that the non-linearity, $\upomega{\times} \bfu$ of (\ref{ns1}),  is fully depleted by grouping $\bfu$ into $\psi$? Now the coefficient functions of $\bfu$ depend only on $\bfx$ and $t$, not on $\bfu$ itself, the resulting parabolic equation becomes linear, somehow covertly, and hence it is much easier to solve. Ultimately, the vorticity will be a function of $\bfx$ and $t$, $\upomega=f(\bfx,t)$; at least, the last expression $\upomega(\bfx,t)$ is always correct in this narrow sense, even without any knowledge of the flow-field. In principle, the whole mathematical endeavour of the governing equations is to determine the functional form $f$ for given initial conditions. The introduction of $\bar{\chi}$, or, $\chi$ (\ref{chi}), has not made any real progress in this respect.
\subsection*{Leray scale in ideal flow}
We have just shown that the quantity (\ref{chi}) is not controlled by the maximum principles of the linear equation. In inviscid flows $\nu=0$, it is reduced to $(\bfv + \sigma ((\bfy - \bfb).\nabla)\chi^* = 0$, which is a senseless identity, as it cannot be true for arbitrary vector $\bfb$ unless $\chi^* {\equiv}0$. To bypass the use of functional $\chi$ and the extremum principles, Chae (2006) assumed certain characters on his vorticity field. There are no {\it a priori} justifications for all of the assumptions. In particularly, he introduced an exponential decay at infinity. Consider his Theorem 1.1 which states, if $ \bfw \in L^1_{loc}(\real^3)$ and there exist positive constants $Y$, $C_0$, $\epsilon_1$, and $\epsilon_2$, such that $|\bfw(\bfy)|\leq C_0 \exp(-\epsilon_1 |\bfy|^{\epsilon_2}) $ for $|\bfy| > Y, \; \forall \bfy  \in \real^3$, then we have $\bfw \in L^{s}(\real^3)$ for $0 < s < 1$. This is a variant of the algebraic decays discussed previously. The law of space translation readily neutralises the apparent benefit of the decay to a less interesting bound $|\bfw(\bfy)| \sim O(C_0)$. A slightly different decay was also assumed in vorticity condition (1.3) and Theorem 1.4 of Chae (2011). 

As stressed by Pomeau (2016), the square-root index in (\ref{leray}) is the only scale of similarity transforms which preserves energy and the constancy of circulation (this is Kelvin's circulation theorem in ideal flows). He treated Chae's work as incomplete though he did not realise the non-existence of the decay. In fact, he was misguided to assume that velocity $\bfv$ may be decreasing like $1/|\bfy|$ at large distances, or vorticity $|\bfw| \sim |\bfy|^{-2}$ for $\bfw \in L^{s}$ as $s \rightarrow 0$. He then argued, that the Leray solutions in such a scenario were unknown. Pomeau went on to suggest the solution of (\ref{le}) may be found by a Laurent expansion, which has the leading order for $|\bfy| \rightarrow \infty$, $\bfv(\bfy) \sim O\big(\hat{\bfv}(\bfy)/|\bfy|\big)$, for some local function $\hat{\bfv}$. Formally, he proposed the expansion
\begin{equation} \label{lt}
\bfv(\bfy) = \sum_{k=1}^{\infty} \frac{\hat{\bfv}_k(\bfy)}{|\bfy|^k}.
\end{equation}
An analogous expansion was applied to the pressure, giving rise to an iterative integral system to determine (non-vanishing) bounded $\hat{\bfv}_k$. Pomeau was unable to provide a complete solution of his integral system. This is hardly surprising, because his arguments overlook the invariant law in $\real^3$. Large distances can be shifted arbitrarily close to the origin, so that the expansion (\ref{lt}) behaves like a divergent series containing an algebraic singularity of infinite order. Beyond doubt, Pomeau's discourse on reviving the Leray conjecture is unfeasible.\footnote{We find it difficult to comprehend the averment that the Leray singularities have been observed and measured in a conventional wind-tunnel, as claimed by Le Berre \& Pomeau (2018). We oblige to reiterate the fact that the blow-up notion is incompatible with the laws of mechanics, and its existence has no theoretical justifications.}

In an attempt to establish a counter-example to the trivial solution of the Leray equations (\ref{leray}), Moffatt (2000; 2009) must have been ill-advised by the purpose-made numerics of Kerr (1993; 1997), and of others. These direct numerical simulations were defective and unreliable, no matter if viscosity is absent or not, as they were carried out in the theoretical framework of periodic domains with periodic boundary conditions, which harbours uncontrolled accumulations of the discretisation errors. In view of the space-invariant principle (\ref{inv}), the proposal to match an `outer' vorticity field with an `inner' blow-up topology, see algebraic expressions (7.11) and (7.12) of Moffatt (2000), is a matter of ambiguity where locality is indefinite. By a simple choice of the vector $\bfb$, the decay of vorticity rebounds into divergence at $\bfy-\bfb \approx 0$, while the interior singularity shrinks to a nought. 
\section{Conclusion}
In summary, the Leray similarity reduction of the Navier-Stokes or incompressible Euler equations, the velocity in (\ref{leray}) and the gradient in (\ref{leray2}), admits only the trivial solution. This is a direct consequence of space homogeneity under the co-ordinates shift involved in Leray's time dilation. In hindsight, the transformed differential equations, (\ref{le}) and 
(\ref{evt}), contain an unspecified coefficient of the co-ordinates, i.e., $\bfy-\bfb$. It must be inconceivable for existence of any structural solution $\bfv=\bfv(\bfy)\neq0$, which readily implies infinite energy.

The impossibility of establishing velocity or vorticity decays at infinity, 
\begin{equation*}
|\bfu|,\;\; |\upomega| \rightarrow |\bfx|^{-k},\;\;\; k > 0,
\end{equation*}
does not imply that these flow quantities are not localised around some clustered core values. Rather, there do not exist {\it a priori} bounds of decays simply because the behaviours of any finite-energy flow depend on the initial data. Should the spatial distribution of the initial velocity be fixed relative to a given translation vector $\bfb$, and in view of the regularity $\upomega \in C^{\infty}([0,\infty))C^{\infty}_0(\real^3)$, the evolved flow quantities do decay at relative distances under the auspices of viscous diffusion, in accordance with the law of energy conservation. 

As discussed earlier, the time-dilation of Leray (\ref{leray}) does not effectively transform the non-linear term $(\bfu.\nabla)\bfu$ of the dynamics. On the face of it, the equations of motion are expressed or re-written into the quasi-steady system (\ref{le}) with the exact non-linearity $(\bfv.\nabla)\bfv$. A mere re-definition of the non-linearity without explicit qualitative formulas can hardly deplete any non-linear effects. This is the reason why the well-established maximum principles for linear equations are not working.

The triviality of Leray's self-similar flows is consistent with the global regularity of incompressible viscous flows established from the vorticity equation where finite-time singularities of any forms are completely ruled out. The quintessence of fluid motions lies in its {\it glocal} shearing interactions, where the non-linear circumstantial evolution from known data proceeds over the space of reciprocality. 

In some of the references quoted above, there are indiscriminate applications of the Sobolev inequality. Appendix A highlights its logical inconsistency. Consequently, the derived ``solutions" of the Navier-Stokes or the Euler equations have limited validity, if not misleading. 
\appendix{A counterexample to the Sobolev inequality} \label{app:a}

The Sobolev inequality plays a fundamental role in the analysis of partial differential equations, see, for example, Adams \& Fournier (2003); Gilbarg \& Trudinger (1998); Majda \& Bertozzi (2002); McOwen (2003); Evans (2008). 

Let $p$, $q$ denote positive real numbers. Consider the Lebesgue spaces $L_p$ in $\real^3$,
\begin{equation} \label{ae1}
\|u\|_{L_p(\real^3)} = \pbrk{ \;\int\limits_{\real^3} \big| u(\bfx) \big|^p \; \rd \bfx }^{\dfrac{1}{p}} \; < \; \infty,
\end{equation}
for all measurable non-zero scalar functions $u$, $\bfx=(x_1,x_2,x_3)$. Functions in $L_p$ are equal almost everywhere; the members of $L_p$ are known as equivalence classes of measurable functions. The important point is that every member of $\|u\|_{L_p}$ is {\it finite} for every given $p$, even though $u$ may become unbounded at specific points in $\real^3$. 

Let $1 \leq p < 3$ and $q=\dfrac{3p}{3-p}$. Suppose that there exists a constant $C=C(p)>0$ such that the following inequality holds,
\begin{equation} \label{ae2}
\|u\|_{L_q(\real^3)} \leq C(p) \; \|D u\|_{L_p(\real^3)},
\end{equation}
for all differentiable scalar functions $u$ with compact support on $\real^3$. Since both $\|u\|_{L_q(\real^3)}$ and $\|D u\|_{L_p(\real^3)}$ can be assigned as finite real numbers, we readily recognise that inequality (\ref{ae2}) is a relative inequality and, thus, suffers from a circular reasoning. Most claims made on the norms of the functions $u$ and its first-order generalised derivatives are imprecise and indefinite, as there exists a finite non-zero constant $1/C^*(p)>0$ such that a reversed inequality of (\ref{ae2}) is also valid. 

Let $u$ denote the function
\begin{equation} \label{ae3}
u=u(\bfx)=\exp \big( - x_1^2 -x_2^2 - x_3^2 \big), \;\;\;\;\; -\infty < x_1,x_2,x_3 < + \infty.
\end{equation}
Take $p=2$ and then $q=6$ in (\ref{ae2}). Direct calculations yield
\begin{equation} \label{ae4}
\|u\|_{L_6(\real^3)} = \bigg(\frac{\pi}{6}\bigg)^{\frac{1}{4}},
\;\;\;\;\;\;
\|D u\|_{L_2(\real^3)} = \sqrt{3} \bigg(\frac{\pi}{2}\bigg)^{\frac{3}{4}}.
\end{equation}
Note that
\begin{equation} \label{ae5}
\pi \bigg(\frac{\pi}{6}\bigg)^{\frac{1}{4}} = \sqrt{\pi} \frac{\pi^{\frac{3}{4}}}{\sqrt[4]{2} \; \sqrt[4]{3}} > \sqrt{\pi} \frac{\pi^{\frac{3}{4}}}{\sqrt[4]{2} \; \sqrt[4]{4}} = \sqrt{\pi}  \bigg(\frac{\pi}{2}\bigg)^{\frac{3}{4}}.
\end{equation}
Thus,
\begin{equation} \tag{$\rm{A}\: 2^{\dagger}$}\label{ae2d}
\|D u\|_{L_2(\real^3)} < C^* \; \|u\|_{L_6(\real^3)}, \;\;\;\; \text{where constant}\: C^* \geq \pi.
\end{equation}
In this counter example to proposition (\ref{ae2}), we interpret $C^*=C^*(C(p))=C^*(p)$. It is straightforward to provide additional examples of interest: (1) $q=2$, $p=6/5$; (2) $p=1$, $q=3/2$, in an exact manner. Evidently, different $C^*$'s are required.

There are no difficulties in generalising our idea to disprove the validity of the Sobolev inequality for vector-valued functions, $\bfu(\bfx,t)$, that are compactly supported on $\real^n$, where, by analogy,
\begin{equation} \tag{$\rm{A}\: 1^*$} \label{ae1s}
\|\bfu(\cdot,t)\|_{L_p(\real^n)} = \pbrk{ \;\int\limits_{\real^n} \big| \bfu(\bfx,t) \big|^p \; \rd \bfx }^{\dfrac{1}{p}} \; < \; \infty,
\end{equation} 
at every given $t < \infty$, and
\begin{equation} \tag{$\rm{A}\: 2^*$}\label{ae2s}
\|\bfu\|_{L_q(\real^n)} \leq C \; \|\nabla \bfu\|_{L_p(\real^n)}.
\end{equation}
It is crucial to take into account any constraints on the vectors, see Lam (2023).

\vspace{1cm}
\begin{acknowledgements}
\noindent 
22 February 2024

\noindent 
\texttt{f.lam11@yahoo.com}
\end{acknowledgements}
\end{document}